\documentclass[%
reprint,
superscriptaddress,
 amsmath,amssymb,
 aps,
pre,
showkeys]{revtex4-2}

\usepackage{graphicx}
\usepackage{dcolumn}
\usepackage{bm}
\usepackage{physics}
\usepackage{footnote}
\usepackage{tabularx}
\usepackage{graphicx}
\usepackage{cleveref}
\usepackage{amsmath}
\usepackage{amssymb}
\usepackage{mhchem}
\usepackage{bbm}
\usepackage[FIGTOPCAP, normalsize]{subfigure}
\usepackage{tikz}
\usepackage{etoolbox}
\usepackage{color}
\providecommand{\selectlanguage}[1]{\relax}
\renewcommand{\selectlanguage}[1]{\relax}
\newcommand{\beq}{\begin{equation}\begin{aligned}}
\newcommand{\eeq}{\end{aligned}\end{equation}}

\Crefname{equation}{Eq.}{Eqs.}
\Crefname{figure}{Fig.}{Figs.}
\crefalias{aligned}{equation}

\usepackage{titlesec}

\begin{document}

\preprint{}

\title{Diffusion with conserved marginal distributions\\
and information theory in fracton hydrodynamics}

\author{Vaibhav Mohanty}\thanks{Correspondence: mohanty@hms.harvard.edu (VM) and sunghanro@fas.harvard.edu (SR)}
\affiliation{Department of Chemistry and Chemical Biology, Harvard University, Cambridge, MA 02138}
\affiliation{Harvard/MIT MD-PhD Program, Harvard Medical School, Boston, MA 02115 and Massachusetts Institute of Technology, Cambridge, MA 02139}
\affiliation{Program in Health Sciences and Technology, Harvard Medical School, Boston, MA 02115 and Massachusetts Institute of Technology, Cambridge, MA 02139}
\author{Sunghan Ro}\thanks{Correspondence: mohanty@hms.harvard.edu (VM) and sunghanro@fas.harvard.edu (SR)}
\affiliation{Department of Physics, Harvard University, Cambridge, MA 02138}


\begin{abstract}
Diffusion with multipole-moment conservation gives rise to transport laws that generalize Fick's law and has attracted growing attention following experimental advances in strongly tilted optical lattices. It was recently shown that conserving complete multipole-moment groups leads to subdiffusive dynamics governed by a nonlinear diffusion equation, raising the question of whether hydrodynamic equations would also be nonlinear when the conservation law is imposed only at the subsystem level. Here we show that subsystem symmetries generically produce nonlinear hydrodynamic equations with shear-only transport, in which any localization present in the initial marginal distributions is preserved at long times by the conservation of those marginals. A linear regime emerges only as a limiting case for small fluctuations around a uniform background. We derive the deterministic and fluctuating parts of the hydrodynamic equations in arbitrary dimensions and obtain the corresponding maximum-entropy equilibrium distributions under constrained marginals. We also show that marginal-conserving diffusion provides a concrete hydrodynamic realization of partial multipole-moment conservation, and we offer an information-theoretic interpretation in which total correlation decays monotonically even when pairwise mutual information does not.

\end{abstract}

\maketitle

\section{INTRODUCTION}
Diffusion is a ubiquitous process in physics that occurs when a locally conserved scalar quantity $\rho({\bf r}, t)$, such as the density, is transported through space via random processes. Due to the conservation condition, the dynamics obey the continuity equation 
\beq \label{eq:cont}
\partial_t \rho ({\bf r}, t) =  - \div \vb{J} ({\bf r}, t) \, .
\eeq
The microscopic picture of diffusion as a random walk leads to the familiar Fick's Law~\cite{tyrrell_origin_1964}. In the hydrodynamic limit of large length and time scales, this yields
\beq
\vb{J} = - D\grad \rho \, ,
\eeq
which, when combined with \Cref{eq:cont}, leads to the diffusion equation 
\beq
\partial_t \rho =  \div(D\grad\rho) \, .
\eeq


Recently, experiments on quantum particles in strongly tilted optical lattices~\cite{guardado2020subdiffusion,scherg2021observing,adler2024observation} have reported subdiffusive dynamics with detailed balance that deviate from the standard Fickian framework~\cite{iaconis2019anomalous,gromov_fracton_2020,iaconis2021multipole,han_scaling_2024,feldmeier20,meerson2024relaxation}. The same constraints have been argued to underlie a wide range of non-ergodic phenomena in tilted lattices and dipole-conserving Hamiltonians, including Stark many-body localization~\cite{schulz2019stark,sachdev02,pielawa11} and Hilbert-space fragmentation~\cite{khemani2020localization,sala2020ergodicity,pollmann20,pai2019localization,moudgalya2021spectral,feng2022hilbert}, and they have motivated a growing literature on the dipolar Bose-Hubbard model and related systems~\cite{lake1,lake2,feldmeier,lake2023non,anakru2023non,will2024realization}. These systems probe the dynamics of fractons~\cite{gromov2024colloquium,pretko2020fracton,pretko17,prem18,pretko18,nandy2024emergent,zerba2025emergent}---quasiparticles that are immobile in isolation but can move collectively---whose thermalization is non-Fickian because the dynamics conserve not only the total particle number but also moments of the particle distribution throughout the diffusion process. For example, if the center of mass
\[
\int \dd[d]{\vb{r}} \vb{r} \rho(\vb{r},t)
\]
is conserved for all times $t$, the density obeys the continuity equation~\cite{gromov_fracton_2020}
\beq
    \partial_t \rho = - \partial_i \partial_j J_{ij},
    \label{eq:ward}
\eeq
where we use the Einstein summation convention. Here, the rank-2 current tensor $J_{ij}$ is the fundamental current, rather than the Fickian current $J_i=\partial_j J_{ij}$. Recent theoretical work in Ref.~\cite{han_scaling_2024} showed that in one dimension this dynamics is described by the nonlinear hydrodynamic equation
\beq \nonumber
    \partial_t \rho = -\partial_x^2 J,
\eeq
with current
\beq \nonumber
    J = D\,\rho^2\,\partial_x^2 \log \rho,
\eeq
where $D$ is a diffusion constant with dimensions $[\mathrm{length}]^5[\mathrm{time}]^{-1}$. The theory predicts localization of particle density at the boundaries of finite systems when the center of mass is conserved, and even bulk localization when higher-order moments are conserved~\cite{han_scaling_2024}.

\begin{figure}[t!]
    \center
    \includegraphics[width=\linewidth]{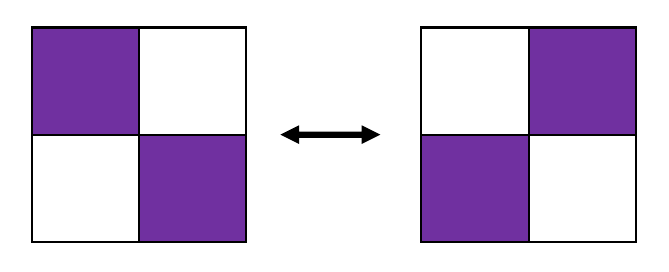}
    \caption{Microscopic model for shear-only transport, where one-dimensional marginal distributions are conserved, and the subsystem symmetries include conservation of mass along every row and column of the lattice. Purple lattice sites are occupied.}
    \label{fig1}
\end{figure}

Ref.~\cite{han_scaling_2024} provides further generalizations to higher-order multipole moment conservation, where complete conservation of all $n$-th degree polynomials $P(\vb{r})$ results in
\beq
    \partial_t \rho = -\partial_A J_A
\eeq
with current
\beq
    J_A = (-1)^{n+1} D\rho^m \partial_A \log \rho,
    \label{eq:general}
\eeq
where $A = \{a_1,\dots,a_{n+1}\}$ is a composite index and $\partial_A = \partial_{a_1}\cdots\partial_{a_{n+1}}$. The value $m \geq n+1$ is an integer whose minimum possible value depends on the minimum number of particles required to conserve a multipole moment~\cite{han_scaling_2024}.

While the results above look comprehensive, many seemingly simple microscopic processes escape this classification. For example, consider the pairwise diffusion process induced by the shearing microscopic moves shown in \Cref{fig1}. Here, a pair of particles in the second and the fourth quadrants moves to the first and the third quadrants and vice versa. Because the move involves two particles symmetrically placed about a common center, conservation of the first moments is automatic, as expected. However, the move also conserves \emph{moments of all orders} for a \emph{subset} of system variables, such as $x$, $x^2$, $x^3$, $\cdots$ and $y$, $y^2$, $y^3$, $\cdots$, but does not in general conserve multivariable moments such as $xy$. As we show below, this constraint has a direct consequence for the equilibrium distribution: while the joint distribution $\rho(x,y)$ evolves with time, the marginal distributions $\rho(x)$ and $\rho(y)$ do not change, and the steady-state distribution factorizes as $\rho^{\mathrm{eq}}(x,y) = \rho(x) \rho(y)$.

We also consider a generalization of the process that involves a shear move for $2^k$ particles in a $k$-dimensional subspace. As we will show, this process conserves $(k-1)$-dimensional marginal distributions of a $d$-dimensional system whenever $k \leq d$. Conservation of marginal distributions of this kind is referred to as a \textit{subsystem symmetry}---namely, conservation of mass along all lower-dimensional lines or (hyper)planes of the system. Fracton hydrodynamics in the presence of such subsystem symmetries has been studied both theoretically~\cite{gromov_fracton_2020,iaconis2019anomalous,osborne2022infinite,grosvenor2021hydrodynamics,guo2022fracton} and through related experimental realizations of constrained dynamics in tilted optical lattices~\cite{guardado2020subdiffusion,scherg2021observing,adler2024observation}.

\begin{figure*}[t!]
    \center
    \includegraphics[width=\linewidth]{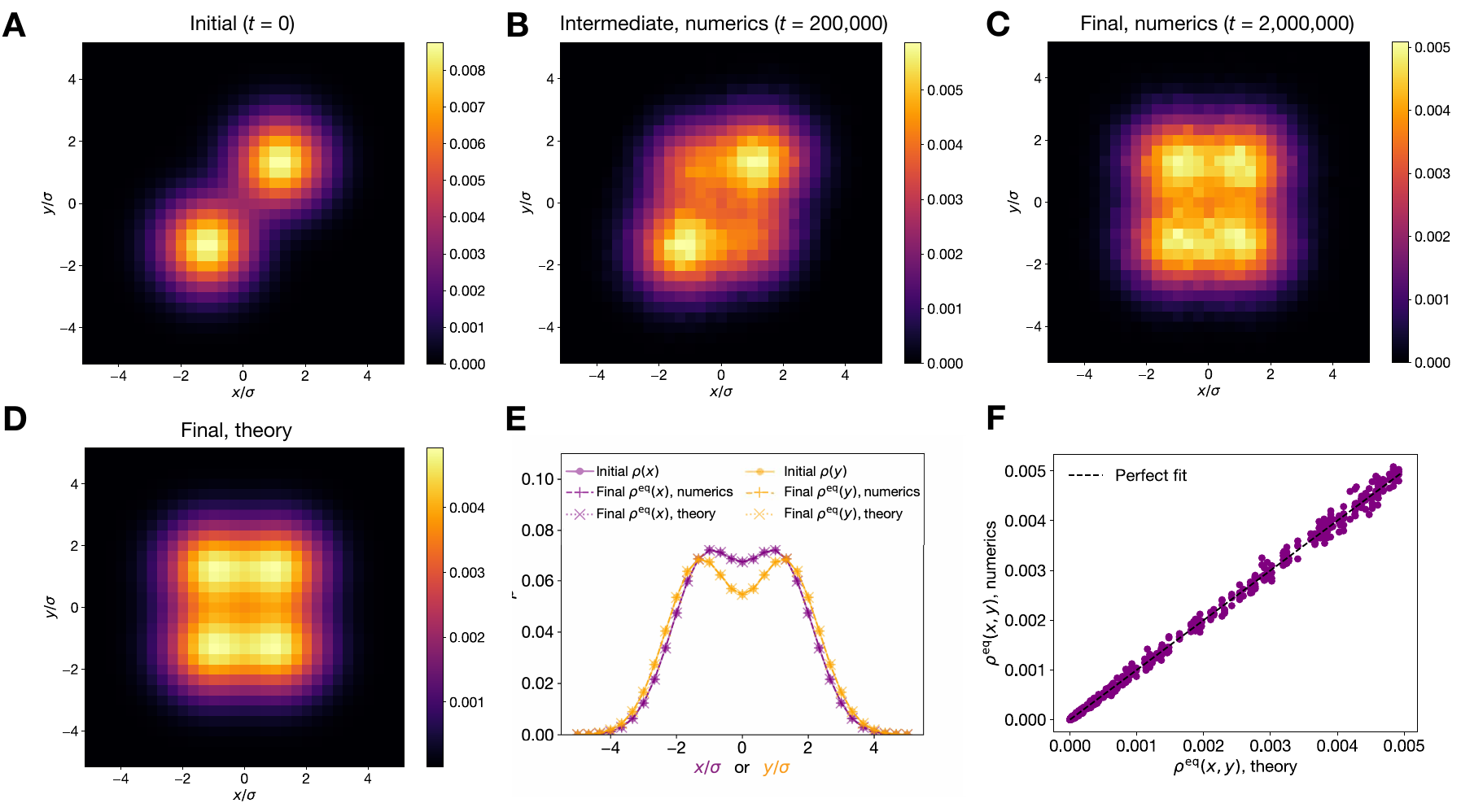}
    \caption{Numerical results for two-dimensional marginal-conserving diffusion. (A) The initial distribution at $t = 0$, (B) an intermediate distribution from numerics at $t = 2 \times 10^5$, (C) the final distribution from numerics at $t = 2 \times 10^6$, (D) the final distribution from theory (i.e. the product distribution). (E) Marginal distributions along each axis for initial, final numerics, and final theory. (F) Agreement (Pearson $r = 0.999$) between $\rho^{\mathrm{eq}}(x,y)$ equilibrium distributions from theory and numerics, where each data point is the density at some point $(x,y)$.}
    \label{fig2}
\end{figure*}

In two dimensions, when mass is conserved along every line of a square lattice, the foundational paper elucidating fracton dynamics in the high-particle-density, hydrodynamic regime---Ref.~\cite{gromov_fracton_2020}---introduces the linearized subdiffusive hydrodynamic equation
\beq
    \partial_t \rho = -C \partial_x^2 \partial_y^2 \rho,
\eeq
with some constant $C$, and generalizes the approach to higher dimensions. In this work, we show that this linear equation appears only as a limiting case for small fluctuations on a uniform background density, and that the hydrodynamic equations valid for arbitrary starting distributions are in fact nonlinear. By introducing a new microscopic model, we analytically derive the general form of the subdiffusive hydrodynamic equations on square and (hyper)cubic lattices in the presence of subsystem symmetries of arbitrary dimension. We find that the nonlinear subdiffusive process, which we validate with numerical simulation, can be realized by shear-only transport in which all diagonal components of the generalized diffusion tensor vanish. The system evolves toward a maximum-entropy distribution with constraints on marginal distributions, for which we obtain an analytical solution. All results are first presented in two dimensions with conservation of one-dimensional marginals and then generalized to arbitrary $d$ dimensions with conservation of $(k-1)$-dimensional marginals. Finally, we introduce an information-theoretic view of fracton hydrodynamics with subsystem symmetries and show that, in $d$ dimensions with $(d-1)$-dimensional subsystem symmetries, the total correlation~\cite{watanabe1960information} between all spatial variables decays monotonically, while the pairwise mutual information between any two coordinates need not.

Our main contributions can be summarized as follows. (i) We construct a microscopic shear model with detailed balance whose continuum limit gives an exact, closed-form nonlinear hydrodynamic equation with conserved marginals in arbitrary dimension. (ii) We obtain the corresponding fluctuating hydrodynamics consistent with the fluctuation-dissipation theorem. (iii) We derive an explicit maximum-entropy form for the equilibrium distribution under arbitrary marginal constraints, generalizing Jaynes' principle~\cite{jaynes1957information}. 
(iv) We resolve an open question raised in Ref.~\cite{han_scaling_2024} regarding the hydrodynamics of partial multipole-moment conservation, showing that only the highest \emph{completely} conserved multipole-moment group governs the dominant dynamics. 
(v) We identify total correlation as a Lyapunov functional for the dynamics, and exhibit a Gaussian counterexample in which pairwise mutual information transiently \emph{increases} along the way to equilibrium.

\section{DIFFUSION WITH SUBSYSTEM SYMMETRIES IN TWO DIMENSIONS}
\label{section_2d}
\subsection{Hydrodynamic equation in two dimensions}
We first introduce a microscopic model for diffusion in two dimensions with conserved one-dimensional marginal distributions. Consider a square lattice. In our microscopic model, we place two particles at opposite corners of a $2\times2$ square (\Cref{fig1}). These particles together may hop to the other (unoccupied) corners of the same square, exchanging the diagonal they occupy. To maintain detailed balance, the forward and reverse moves occur at the same rate $r$. The time evolution of the particle density $\rho_{x,y}(t)$ is given by the master equation
\beq
    \partial_t \rho_{x,y} &= r\left[\rho_{x+1,y} \rho_{x,y+1} + \rho_{x-1,y} \rho_{x,y+1}\right. \\
    &\qquad\left. + \rho_{x,y-1}\rho_{x+1,y} + \rho_{x,y-1} \rho_{x-1,y} \right. \\
    &\left.- \rho_{x,y} \left(\rho_{x+1,y+1} + \rho_{x-1,y+1} \right.\right.\\
    &\qquad\left.\left.+ \rho_{x+1,y-1} + \rho_{x-1,y-1}\right)\right].
\eeq
Expanding to leading non-vanishing order in the lattice constant $a$ (subsequently set to $a = 1$), we obtain the hydrodynamic limit
\beq
    \partial_t \rho({\bf r},t) &= - \partial_i\partial_j J_{ij}({\bf r},t) \,,
\label{eq:2d_shear}
\eeq
with current tensor
\beq
    J_{ij} = D_{ijkl} \, \partial_k \partial_l \log \rho + \mathcal{O}(\partial^4) \, ,
\label{eq:2d_current}
\eeq
where $D_{ijkl}$ is a generalized diffusion tensor with dimensions $[\mathrm{length}]^2\,[\mathrm{time}]^{-1}$. The pairwise structure of the microscopic move and detailed balance implies that $J_{ij}=J_{ji}$ which is satisfied through the diffusion tensor index symmetries
\beq
    D_{ijkl} = D_{jikl} = D_{klij} \, .
\eeq
The explicit form derived from the microscopic process is
\beq
    D_{ijkl} = \frac{D \rho^2 }{4} (1 - \delta_{ij}) (1 - \delta_{kl}) ( \delta_{ik} \delta_{jl} + \delta_{il} \delta_{jk}) \, ,
\eeq
in which the factor $\rho^2$ reflects that the elementary move involves two particles. Substituting this into Eqs.~\eqref{eq:2d_shear}-\eqref{eq:2d_current} yields the closed-form deterministic equation
\beq
    \partial_t \rho = - D \, \partial_x \partial_y \left[ \rho^2 \, \partial_x \partial_y \log \rho \right] \, .
    \label{eq:2d_master_hydro}
\eeq

Notice that this is consistent with the structure of \Cref{eq:ward}, with the additional property that the diagonal currents $J_{xx}$ and $J_{yy}$ vanish while only the off-diagonal current $J \equiv J_{xy} = J_{yx}$ is non-zero, signalling shear-only transport. The $\mathcal{O}(\partial^4)$ terms omitted in \Cref{eq:2d_current} are subleading in the long-wavelength limit and do not affect any of the conclusions below.

We claim that \Cref{eq:2d_master_hydro} is the nonlinear extension of the equation introduced in Ref.~\cite{gromov_fracton_2020} to describe diffusion with conserved particle density along every row and column of the square lattice. The microscopic model makes it immediately apparent that, because particles alternate lattice sites, particle density does not change along any row or column after any microscopic step. Mathematically,  marginal distribution conservation is also observed from the hydrodynamic equation directly by integrating over either $x$ or $y$:
\beq
    \partial_t \rho(x,t) = \int \dd{y} \partial_t \rho(x,y,t) = -D \partial_x \int \dd{y} \partial_y J,
\eeq
which is a vanishing boundary term. In terms of moment conservation, it immediately follows that any monomial $x^n$ or $y^n$ which does not contain cross products such as $xy$ will be conserved:
\beq
    \partial_t \int \dd{x} \int \dd{y} x^n \rho(x,y,t) &= \partial_t \int \dd{x} x^n \rho(x,t) \\
    &= \partial_t \int \dd{x} x^n \rho(x) = 0,
\eeq
with the same applying if we replace $x$ with $y$. 

The linearized equation of Ref.~\cite{gromov_fracton_2020} is recovered in the limit where the density field $\rho$ consists of a uniform background $\rho_0$ plus a small fluctuation $\delta \rho$. Writing $\rho = \rho_0 + \delta \rho$,
\beq
    \partial_t \rho &= -D \, \partial_x\partial_y \left[(\rho_0 + \delta \rho)^2 \, \partial_x\partial_y \log(\rho_0 + \delta \rho)\right],
\eeq
and expanding to leading order in $\delta \rho / \rho_0$ gives
\beq
    \partial_t \delta \rho  \approx - \widetilde{D} \, \partial_x^2 \partial_y^2 \, \delta \rho,
\eeq
where $\widetilde{D} \equiv D \rho_0$ is the rescaled diffusion constant. Thus \Cref{eq:2d_master_hydro} is the more general hydrodynamic equation with conserved one-dimensional marginal distributions.

\subsection{Fluctuating hydrodynamics by thermal noise}

The deterministic hydrodynamic equation \Cref{eq:2d_master_hydro} applies in the limit of vanishing thermal noise. In the presence of thermal noise, the current acquires an additional fluctuating contribution $\xi_{ij}$:
\beq
    J_{ij} ({\bf r}, t) = D_{ijkl} \, \partial_k \partial_l \log \rho + \xi_{ij}({\bf r}, t)\, .
\eeq
The noise correlation can be fixed by requiring that the system relaxes to thermodynamic equilibrium without steady-state entropy production. In the case $\xi_{ij}({\bf r}, t)$ is a Gaussian white noise, following the Martin-Siggia-Rose-Janssen-de Dominicis approach~\cite{martin1973statistical,janssen1976lagrangean,lefevre2007dynamics} detailed in \Cref{appen:noise}, one obtains
\beq
    \langle \xi_{ij}({\bf r},t) \xi_{kl}({\bf r}',t') \rangle
    = 2 D_{ijkl} \, \delta({\bf r}-{\bf r}') \, \delta(t-t') \, ,
    \label{eq:2d_fluc}
\eeq
together with a generalized Einstein relation $D_{ijkl} = \mu_{ijkl} k_{\rm B} T$ between the diffusion tensor and the underlying mobility tensor $\mu_{ijkl}$. The correlator in \Cref{eq:2d_fluc} is therefore consistent with the fluctuation-dissipation theorem.

\subsection{Maximum entropy and the equilibrium distribution}

\begin{figure*}[t!]
    \center
    \includegraphics[width=0.7\linewidth]{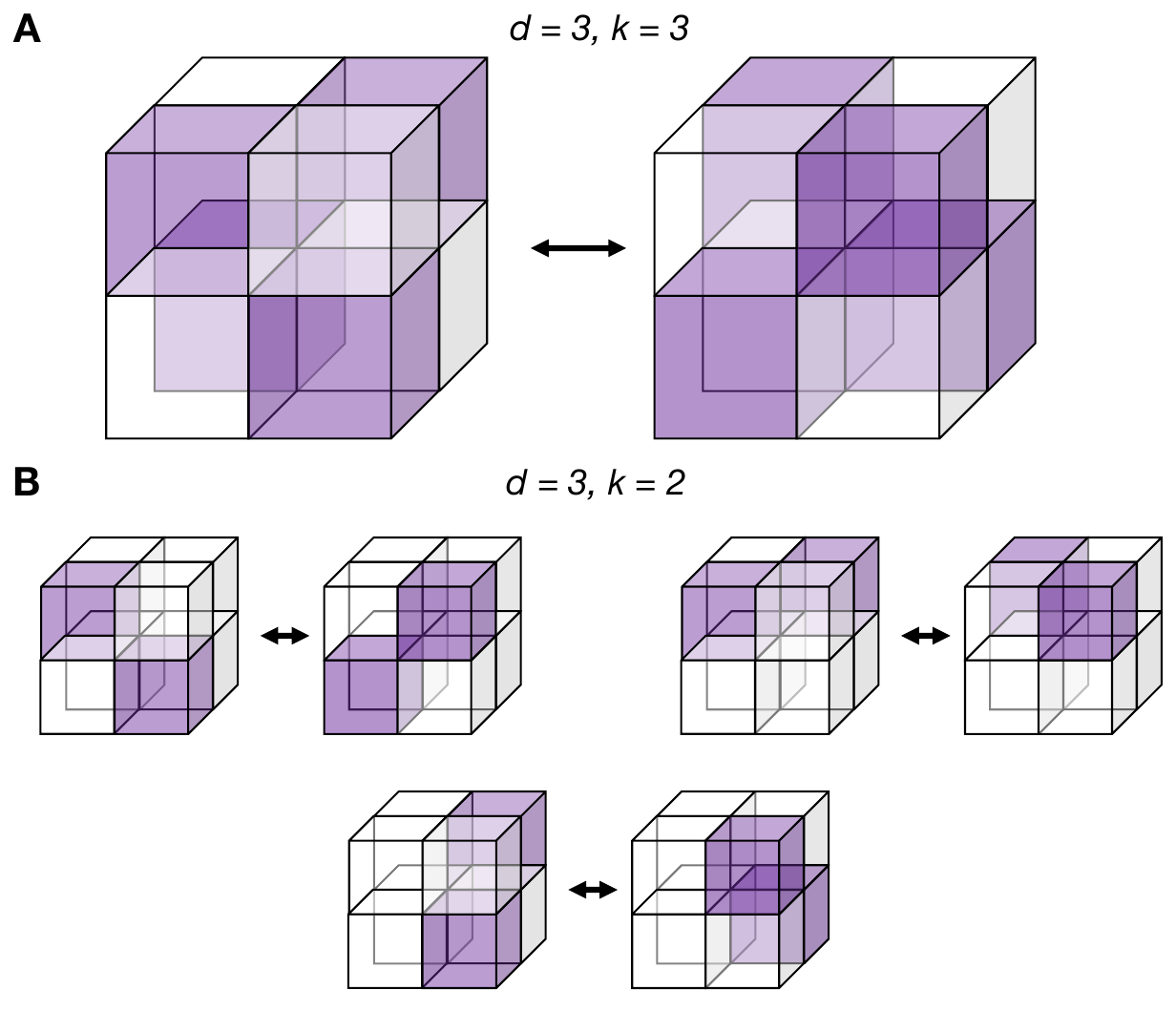}
    \caption{Generalized microscopic model for (A) $d=3$, $k=3$, where two-dimensional (and one-dimensional) marginal distributions are conserved, and the associated subsystem symmetries include conservation of mass along every line and every plane; and for (B) $d=3$, $k=2$, where only one-dimensional marginal distributions are conserved, and the associated subsystem symmetries include conservation of mass along every plane but not every line. Purple lattice sites are occupied.}
    \label{fig3}
\end{figure*}

We now study the equilibrium distribution by maximizing the Shannon-Jaynes entropy under the appropriate constraints, following Refs.~\cite{jaynes1957information,han_scaling_2024,csiszar1975divergence}. Throughout we work directly with the single-particle density $\rho({\bf r}) = \langle \hat{\rho}({\bf r}) \rangle$, where the empirical density is $\hat{\rho}({\bf r}) \equiv \sum_{i=1}^N \delta({\bf r} - {\bf r}_i)$ and the average is taken with respect to the many-body distribution; this is justified because the relevant constraints in the master equation (number conservation and conservation of marginals) are linear in $\rho({\bf r})$.

In two dimensions, the maximum-entropy distribution constrained by both one-dimensional marginals is simply the product of those marginals (see Appendix~\ref{MaxEnt_2D}):
\beq
    \rho^{\mathrm{eq}}(x,y) = \rho(x)\rho(y).
    \label{eq:2d_maxent_sol}
\eeq
Thus, the equilibrium distribution $\rho^{\mathrm{eq}}(x,y)$ is given by the product of the marginals, which are determined by the initial distribution. As a consistency check, if we substitute \Cref{eq:2d_maxent_sol} into the current in \Cref{eq:2d_current}, we find that the current vanishes
\beq
    J &= \rho^2\partial_x \partial_y \log \rho(x)\rho(y) \\
    &= \rho^2\partial_x \partial_y \left[\log \rho(x) + \log\rho(y)\right] = 0.
\eeq
Therefore, the system is at equilibrium.

We verify the theoretical equilibrium distribution by performing numerical simulations on the square $[-10,10]^2$, discretized as a lattice with $31 \times 31$ sites, with $2 \times 10^5$ particles. The initial distribution is a mixture of two isotropic Gaussians with $\sigma_x = \sigma_y = 3$, centered at $(-3.5, -4)$ and $(3.5, 4)$ (\Cref{fig2}A). While we use hard-wall boundary conditions, since the distributions we study are localized inside the box, the choice of boundary condition does not affect the bulk evolution. 
At each time step, up to $1000$ particle swaps of the type defined by our microscopic model (\Cref{fig1}) are attempted, with each attempt requiring two compatible occupied corners that are randomly chosen. An intermediate distribution after $2\times10^5$ steps is plotted in \Cref{fig2}B, the equilibrated distribution after $2\times10^6$ steps is shown in \Cref{fig2}C, and the theoretical equilibrium distribution from \Cref{eq:2d_maxent_sol} is shown in \Cref{fig2}D. The marginals along both axes at the initial and final times are shown in \Cref{fig2}E. The marginal distributions are preserved throughout the simulation, and the theoretical equilibrium distribution, given by the product of the (initially determined) marginals, agrees closely with the numerically simulated final distribution (\Cref{fig2}F).

\section{GENERALIZED HYDRODYNAMIC THEORY IN ARBITRARY DIMENSIONS}

We now generalize the microscopic process to higher dimensions. In $d$ spatial dimensions, consider a block consisting of $2^{k}$ lattice sites ($2\times 2\times \cdots \times 2$ a total of $k$ times), with $2 \leq k \leq d$. We let alternating lattice sites to contain a particle in the starting configuration. Mathematically, every lattice site can have an index given by the binary sequence $\vb{b} = (b_1,\dots,b_k) \in \{0,1\}^k$, which allows to define the site's parity $\Pi(\vb{b})$
\beq
    \Pi(\vb{b}) = \prod_{i=1}^k(2b_i - 1).
\eeq
From the above definition, we can then define two configurations: in the ``even'' configuration, every site which has positive parity $\Pi(\vb{b}) = +1$ contains a particle, and in the ``odd'' configuration, every site which has negative parity $\Pi(\vb{b}) = -1$ contains a particle. We then define the microscopic process in which even-odd parity switches happen at some rate $r$. An example is shown for $k = d = 3$ in \Cref{fig3}A. If $k < d$, then there will be ${d \choose{k}}$ parallel processes, one for each $k$-dimensional subset of the total $d$ dimensions in the system, as exemplified by \Cref{fig3}B where we have $d = 3$ and $k = 2$.


When $k = d$ (e.g.\ \Cref{fig3}A), the hydrodynamic equation contains a single term with $k = d$ derivatives. When $k < d$ (e.g.\ \Cref{fig3}B), there are ${d \choose{k}}$ terms, each with $k$ derivatives, corresponding to the ${d \choose{k}}$ parallel processes. In general, the dynamics follow
\beq
    \partial_t \rho = -\sum_{\substack{S\subseteq \{1,\dots,d\}\\\abs{S}=k}} \left(\prod_{s \in S}\partial_{s}\right) J_S,
    \label{eq:gendiff}
\eeq
with current
\beq
    J_S = (-1)^{k} D \, \rho^{2^{k-1}} \left(\prod_{s \in S}\partial_{s}\right)\log \rho,
    \label{eq:gencurr}
\eeq
where $S \subseteq \{1,\dots,d\}$ is a subset of spatial indices with cardinality $\abs{S} = k$. The exponent $\rho^{2^{k-1}}$ in the current is the number of particles taking part in a single elementary move. In detail, the microscopic move flips $2^{k-1}$ ``even-parity'' occupations into $2^{k-1}$ ``odd-parity'' occupations within the $2^k$-site block, so the rate of any local move is proportional to the product of $2^{k-1}$ density factors evaluated on neighbouring sites; the gradient expansion of this product yields an effective coefficient $\propto \rho^{2^{k-1}}$ in the hydrodynamic limit. The case $k=2$ recovers $\rho^2$ in agreement with \Cref{eq:2d_master_hydro}.

The general hydrodynamic equations above conserve all marginal distributions of up to dimension $k-1$. The equilibrium distribution is given by maximum entropy with constraints on all $(k-1)$-dimensional marginal distributions chosen over $d$ dimensions~\cite{jaynes1957information}. The general form of this distribution, derived in Appendix \ref{MaxEnt_general}, is
\beq
\rho^{\mathrm{eq}}(\vb{r}) &= \frac{1}{Z}\exp\left[\sum_{\substack{S\subseteq \{1,\dots,d\}\\\abs{S}=k-1}} \phi_{S}(\vb{r}_S)\right], \\
&=\frac{1}{Z}\prod_{\substack{S\subseteq \{1,\dots,d\}\\\abs{S}=k-1}}\exp\left[\phi_{S}(\vb{r}_S)\right]
\label{eq:gensol}
\eeq
where $S \subseteq \{1,\dots,d\}$ is a subset of spatial indices with cardinality $\abs{S} = k-1$, $\vb{r}_S$ is a $(k-1)$-dimensional subset of the $d$-dimensional position vector $\vb{r}$, all of the $\{\phi_{S}(\vb{r}_S)\}$ are functions which must be jointly learned in order to correctly reproduce all $(k-1)$-dimensional marginals, and $Z$ is a normalization constant. There are a number of algorithmic and machine learning methods which can be used to learn the functions $\{\phi_{S}(\vb{r}_S)\}$~\cite{johnson_learning_2007}. Note that substituting \Cref{eq:gensol} into \Cref{eq:gencurr} yields zero current, as expected, because fewer than $k$ unique variables are present in each $\phi_{S}(\vb{r}_S)$.

\begin{figure}[t!]
    \center
    \includegraphics[width=\linewidth]{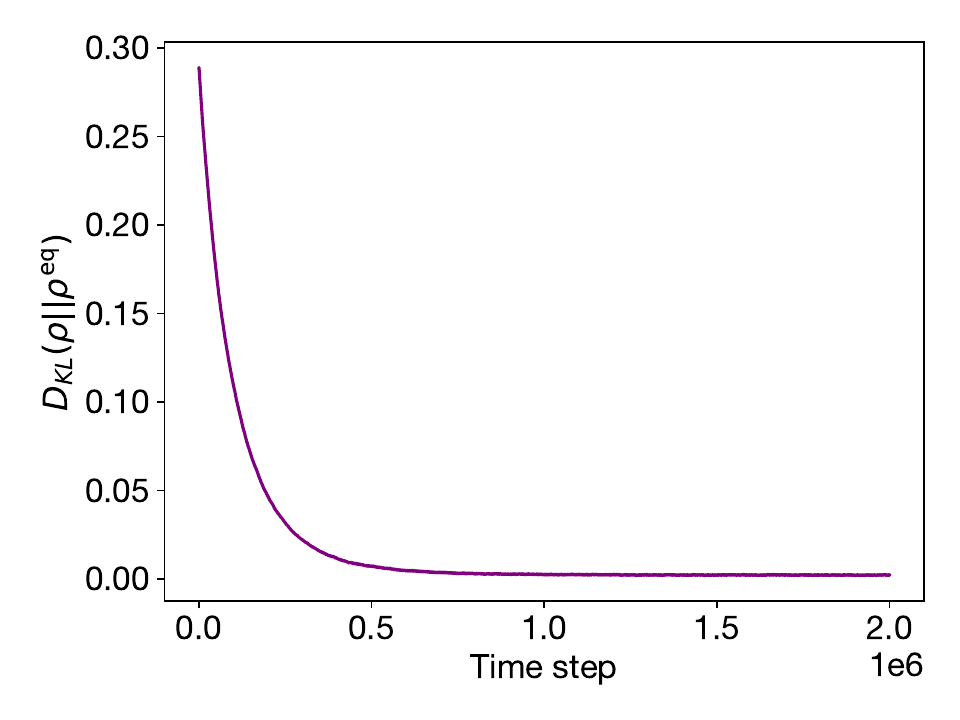}
    \caption{KL divergence $D_{\mathrm{KL}}[\rho||\rho^{\mathrm{eq}}]$ for two-dimensional marginal-conserving diffusion plotted as a function of time, showing monotonic decrease towards zero.}
    \label{fig4}
\end{figure}

We now remark on scaling in the linearized regime. Conservation of the $(k-1)$-dimensional marginals implies that any localization present in the initial marginals persists at equilibrium, so a single-parameter scaling Ansatz for the full density profile is not possible in general. However, as in the two-dimensional case, one can study \Cref{eq:gendiff} in the regime where $\rho$ consists of small fluctuations over a uniform background. Writing $\rho = \rho_0 + \phi$ with $|\phi| \ll \rho_0$ and expanding to leading order recovers the linearized hydrodynamic form from Ref.~\cite{gromov_fracton_2020}:
\beq
    \partial_t \phi = (-1)^{k+1} \widetilde{D} \sum_{\substack{S\subseteq \{1,\dots,d\}\\\abs{S}=k}} \left(\prod_{s \in S}\partial_{s}^2\right) \phi,
\eeq
where $\widetilde{D} \equiv D \, \rho_0^{2^{k-1}-1}$. The associated dynamical scaling exponent for the relaxation of small fluctuations is $z = 2k$.

\section{RELATION TO PARTIAL MULTIPOLE MOMENT CONSERVATION}
Our findings also address an open question raised in Ref.~\cite{han_scaling_2024}, an active topic of investigation in fracton physics. The authors of Ref.~\cite{han_scaling_2024} developed a hydrodynamic theory for diffusion when complete multipole-moment groups of fixed order are conserved, which means that all moments constructed from $n$-th order polynomials are conserved. For instance, in two-dimensional quadrupole conservation, one would have $\int \dd\vb{r}\, \rho(\vb{r}) x^2$, $\int \dd\vb{r}\, \rho(\vb{r}) y^2$, and $\int \dd\vb{r}\, \rho(\vb{r}) xy$ all conserved. Reference~\cite{han_scaling_2024} pointed out that the case of \emph{partial} multipole-moment conservation, in which (for example) $\int \dd\vb{r}\, \rho(\vb{r})  x^2$ and $\int \dd\vb{r}\, \rho(\vb{r}) y^2$ are conserved but $\int \dd\vb{r}\, \rho(\vb{r}) xy$ is not, was left open.

We point out that diffusion with $(k-1)$-dimensional marginals conserved realizes exactly this scenario. In two dimensions, the subsystem symmetry of mass conservation along every row and column of the square lattice is equivalent to conservation of the one-dimensional marginals, in which the moments $\int \dd\vb{r}\, \rho(\vb{r}) x^2$ and $\int \dd\vb{r}\, \rho(\vb{r}) y^2$---but not $\int \dd\vb{r}\, \rho(\vb{r}) xy$---are conserved. More generally, conservation of one-dimensional marginal distributions implies that \textit{all} single-variable moments $x^n$ and $y^n$ for arbitrary $n$ are conserved, while \textit{none} of the mixed-variable moments $x^m y^n$ with $m,n \geq 1$ are conserved.

Reference~\cite{han_scaling_2024} showed that conservation of \textit{all} multipole moments constructed from $(k-1)$-th order polynomials (but not necessarily of higher orders) yields a hydrodynamic equation with $k$ spatial derivatives acting on the current and $k$ spatial derivatives inside the current. In marginal-conserving diffusion with $(k-1)$-dimensional marginals conserved, multipole moments built from $k$-th order polynomials are \textit{completely} conserved, while (infinitely) many higher-order moments are \textit{partially} conserved. We find that the differential operator acting on the current still has $k$ derivatives, one for each group of $k$ variables chosen from $d$ spatial dimensions, summed over all ${d \choose k}$ such groups. Marginal-conserving diffusion therefore demonstrates that the number of spatial derivatives in the hydrodynamic equations is set by the highest order of multipole-moment group that is \textit{completely} conserved; partially conserved multipole-moment groups do not affect the dominant dynamics.

\section{MARGINAL-CONSERVING DIFFUSION AND INFORMATION THEORY}
\subsection{Shannon-Jaynes entropy and KL divergence}
For the generalized processes described in the previous section, the Shannon(-Jaynes) entropy $\mathcal{S} = -\int \dd[d]{\vb{r}} \rho \log (\rho/q)$, with a dimensionful constant $q$, monotonically increases (see Appendix \ref{lyapunov_entropy}):
\beq
    \dv{\mathcal{S}[\rho]}{t} \geq 0.
\eeq
It is also informative to consider the KL divergence between the time-dependent particle density distribution and the equilibrium distribution
\beq
    D_{\mathrm{KL}}[\rho||\rho^{\mathrm{eq}}] &\equiv \int \dd[d]{\vb{r}} \rho(\vb{r}) \log \frac{\rho(\vb{r})}{\rho^{\mathrm{eq}}(\vb{r})} = \mathcal{S}[\rho | \rho^{\mathrm{eq}}] -  \mathcal{S}[\rho],
\eeq
where $\mathcal{S}[\rho | \rho^{\mathrm{eq}}]$ is the cross-entropy between $\rho$ and $\rho^{\mathrm{eq}}$. While the Shannon entropy's asymptotic value is not necessarily easy to calculate (because it requires calculating the equilibrium distribution), the KL divergence is guaranteed to approach zero asymptotically and thus may be a more interpretable metric for quantifying the approach to equilibrium. The cross-entropy between the particle distribution $\rho$ and the maximum entropy distribution $\rho^{\mathrm{eq}}$ is exactly equal to the Shannon entropy of $\rho^{\mathrm{eq}}$ in equilibrium:
\beq
    \mathcal{S}[\rho|\rho^{\mathrm{eq}}] = \mathcal{S}[\rho^{\mathrm{eq}}],
\eeq
which means that the KL divergence is exactly equal to the entropy difference between $\rho$ and $\rho^{\mathrm{eq}}$:
\beq
    D_{\mathrm{KL}}[\rho||\rho^{\mathrm{eq}}] = \mathcal{S}[\rho^{\mathrm{eq}}] -  \mathcal{S}[\rho].
\eeq
In general, this is true for distributions $\rho$ which obey the same constraints placed on the maximum entropy distribution $\rho^{\mathrm{eq}}$; we prove it explicitly for marginal-conserving diffusion in Appendix \ref{lyapunov_kldiv}. Now, taking a time derivative, it immediately follows that the KL divergence also decreases monotonically with time:
\beq
    \dv{D_{\mathrm{KL}}[\rho||\rho^{\mathrm{eq}}]}{t} = -\dv{\mathcal{S}[\rho]}{t} \leq 0.
\eeq
For the two-dimensional numerical simulation studied in Section \ref{section_2d}, we plot KL divergence over time to verify its monotonicity and approach to zero in \Cref{fig4}. 

\subsection{Mutual information and total correlation}

The central information-theoretic result of this work is the following. In $d$ dimensions with $(d-1)$-dimensional subsystem symmetry (i.e.\ $k=2$), the total correlation~\cite{watanabe1960information} between all $d$ spatial coordinates of a particle is a Lyapunov functional of the dynamics, decreasing monotonically to zero. The pairwise mutual information between any two of those coordinates, however, is generally \emph{not} monotone, and we exhibit an explicit counterexample below. The two-dimensional case is special because total correlation and pairwise mutual information coincide.

In the two-dimensional case ($d = k = 2$), suppose we randomly sample a position vector $(X(t),Y(t))$ according to the density $\rho(\vb{r},t)$ at finite time $t$, and sample another position vector $(X(t\rightarrow\infty),Y(t\rightarrow\infty))$ from the equilibrium distribution $\rho^{\mathrm{eq}}(\vb{r})$. The mutual information between any two spatial components, with one chosen at \textit{any} finite time $t$ and the other chosen at equilibrium ($t\rightarrow \infty$), is always zero
\beq
    I[X(t),X(t\rightarrow \infty)] &= 0 \\
    I[Y(t),Y(t\rightarrow \infty)] &= 0 \\
    I[X(t),Y(t\rightarrow \infty)] &= 0 \\
    I[Y(t),X(t\rightarrow \infty)] &= 0.
\eeq
because $(X(t\rightarrow\infty),Y(t\rightarrow\infty))$ is drawn independently from the equilibrium distribution regardless of $(X(t),Y(t))$. Although the equilibrium and finite-time distributions are related, a particle with known position sampled at time $t = 0$ will be found at long times according to the equilibrium distribution $\rho^{\mathrm{eq}}(\vb{r})$ without any dependence on the initial position. This is a well-understood result that applies to any ergodic diffusion process, not just marginal-conserving diffusion.

But, in marginal-conserving diffusion, what happens to the mutual information \textit{between spatial coordinates} at \textit{any given (equal) time}? That is, how does the equal-time mutual information
\beq
    I[X(t),Y(t)] = \int \dd[d]{\vb{r}} \rho(x,y,t) \log \frac{\rho(x,y,t)}{\rho(x,t)\rho(y,t)}
\eeq
behave as a function of $t$? We first note that the marginal distributions $\rho(x,t) = \rho(x)$ and $\rho(y,t) = \rho(y)$ are time-independent, since they are conserved by the process. It is then clear that the mutual information between the two spatial coordinates is exactly the KL divergence between the particle distribution at time $t$ and the equilibrium distribution, which is the product distribution in two dimensions (\Cref{eq:2d_maxent_sol}):
\beq
    I[X(t),Y(t)] = D_{\mathrm{KL}}[\rho(x,y,t)||\rho^{\mathrm{eq}}].
\eeq
Thus, importing the results for $D_{\mathrm{KL}}[\rho||\rho^{\mathrm{eq}}]$, it immediately follows that $I[X(t),Y(t)]$ also decays monotonically to zero over the course of the diffusion process. Thus, marginal-conserving diffusion in two dimensions \textit{dissipates mutual information between spatial coordinates} monotonically over time. This is not necessarily true for other diffusion processes, or even when $d > 2$ with $k > 2$.

The $d > 2$ and $k = 2$ cases are where the underlying subsystem symmetries are the conservation of mass density within every $(d-1)$-dimensional hyperplane orthogonal to one of the Cartesian axes. Here, the only conserved marginal distributions are one-dimensional marginals, so the equilibrium distribution is a product distribution
\beq
    \rho^{\mathrm{eq}}(\vb{r}) = \prod_{i=1}^d \rho(r_i).
\eeq
The KL divergence between $\rho$ and the product distribution above is the \textit{total correlation}~\cite{watanabe1960information} between the spatial components of a random particle at position $\vb{R}(t)$ sampled from $\rho(\vb{r},t)$ at time $t$:
\beq
    TC[\vb{R}(t)] = \int \dd[d]{\vb{r}} \rho(\vb{r},t) \log \frac{\rho(\vb{r},t)}{\prod_{i=1}^d \rho(r_i)}.
\eeq
Thus, for $d > 2$ and $k = 2$, the total correlation between all spatial variables is monotonically dissipated toward zero at equilibrium. Moreover, the mutual information between any two spatial variables goes to zero at equilibrium:
\beq
    I[&R_i(t\rightarrow \infty), R_j(t\rightarrow \infty)] \\
    &= \int \dd{r_i} \int \dd{r_j} \rho^{\mathrm{eq}}(r_i,r_j) \log \frac{\rho^{\mathrm{eq}}(r_i,r_j)}{\rho^{\mathrm{eq}}(r_i)\rho^{\mathrm{eq}}(r_j)} = 0,
\eeq
since $\rho^{\mathrm{eq}}(r_i,r_j) = \rho^{\mathrm{eq}}(r_i)\rho^{\mathrm{eq}}(r_j)$.

However, the mutual information between a pair of two variables does \textit{not} necessarily decrease monotonically toward zero at all finite times, despite the total correlation monotonically decreasing. To illustrate this concretely, we construct a specific example with $d = 3$ and $k = 2$. We want to construct an example where the mutual information between two of the three variables, $I[X(t),Y(t)]$ (which goes to zero at long times), does so \textit{non-monotonically}. Thus, there must be some time $t$ where
\beq
    \dv{I[X(t),Y(t)]}{t} > 0.
    \label{eq:nonmonotonicity_condition}
\eeq
At time $t = 0$, suppose we have a three-dimensional Gaussian centered at the origin with a covariance matrix
\beq
    \Sigma = \begin{pmatrix}
    1 & \sigma_{xy} & \sigma_{xz} \\
    \sigma_{xy} & 1 & \sigma_{yz}\\
    \sigma_{xz} & \sigma_{yz} & 1 \\
    \end{pmatrix},
    \label{eq:covariance_mat}
\eeq
with correlations $\sigma_{xy}$, $\sigma_{yz}$, and $\sigma_{xz}$, chosen so that the covariance matrix is appropriately positive semi-definite. At equilibrium, the resulting distribution is an isotropic Gaussian with an identity covariance matrix. In Appendix \ref{gaussian_appendix}, we prove that if the general condition
\beq
    \sigma_{xy}(\sigma_{xy} - \sigma_{xz}\sigma_{yz}) < 0
    \label{eq:final_condition}
\eeq
is met, then at time $t = 0$, the mutual information between $X(t=0)$ and $Y(t=0)$ is increasing 
\beq
    \dv{I[X(t=0),Y(t=0)]}{t} > 0,
\eeq
even if the total correlation between $X(t=0)$, $Y(t=0)$, and $Z(t=0)$ is decreasing. In general, we have shown that the diffusion with $(d-1)$-dimensional subsystem symmetries \textit{dissipates total correlation} between spatial variables, even when mutual information between pairs of variables is not decreasing. There may be experimental realization of fracton hydrodynamics which can exhibit this decorrelation.

\section{DISCUSSION}
In this work, we studied diffusion constrained by subsystem symmetries, such as conservation of mass along lines or (hyper)planes of a lattice. The microscopic model, which is designed to exhibit shear-only transport, yields nonlinear hydrodynamic equations in which each differential operator carries only a single spatial derivative along each of $k$ axes in the current. The current therefore vanishes whenever the particle density distribution factorizes into pieces depending on at most $k-1$ variables each. We then derived the maximum-entropy distribution constrained by all $(k-1)$-dimensional marginals. Our theory extends the linear hydrodynamic equations of Ref.~\cite{gromov_fracton_2020}, which we recover in the limit of small fluctuations on a uniform background. In this regime, the approach to equilibrium is characterized by a dynamical scaling exponent $z = 2k$.

Next, we showed how marginal-conserving diffusion maps onto, and resolves, the open problem of diffusion with partial multipole-moment-group conservation. Because all marginals of fixed dimension are conserved, there is a ceiling on the order of the multipole-moment group that is completely conserved, even when infinitely many higher orders of polynomials are partially conserved. We found that the dominant dynamics depend only on the highest order of multipole-moment group that is completely conserved; partial conservation does not affect the leading subdiffusive behavior.

Lastly, we developed an information-theoretic perspective on marginal-conserving diffusion. We showed that the Shannon-Jaynes entropy of the particle distribution and the KL divergence between the particle distribution and the equilibrium distribution both evolve monotonically with the dynamics, and that the KL divergence is exactly equal to the entropy gap between the time-dependent and equilibrium distributions. When the subsystem symmetries amount to conservation of mass along every $(d-1)$-dimensional hyperplane orthogonal to a Cartesian axis (equivalently, conservation of all one-dimensional marginals), the diffusion process monotonically dissipates the total correlation~\cite{watanabe1960information} between spatial variables on its approach to equilibrium, where the steady-state distribution is simply the product distribution. However, the mutual information between any chosen pair of spatial coordinates does not necessarily decay monotonically; this asymmetry is invisible in two dimensions ($d = 2$), where mutual information and total correlation coincide.

These information-theoretic statements suggest concrete observables for future experiments. The total correlation can be reconstructed from full-counting statistics of joint occupation numbers in site-resolved quantum gas microscopes of the kind used in Refs.~\cite{guardado2020subdiffusion,scherg2021observing,adler2024observation}, while pairwise mutual information requires only the two-coordinate marginals. Because the dominant dynamical exponent of the relaxation is $z=2k$, the predicted non-monotonic transient in the pairwise mutual information should be visible on a timescale $\tau \sim L^{2k}/\widetilde{D}$ for a system of linear size $L$, set by the slowest relaxing modes of the linearized theory. Realizations of fracton phases of matter in the hydrodynamic regime are now being achieved with diffusion under complete multipole-moment conservation~\cite{guardado2020subdiffusion,scherg2021observing,adler2024observation}, and analogous experiments with engineered subsystem symmetries would provide a direct test of the predictions developed here. Beyond the specific predictions of this work, our results clarify the structural link between information-theoretic functionals and the dynamics of fracton phases of matter.

\section{ACKNOWLEDGMENTS}
This work was supported by a Hertz Foundation Fellowship (to V.M.), a PD Soros Fellowship (to V.M.), and by award T32GM14427 from the National Institute of General Medical Sciences (to Harvard/MIT MD-PhD Program). The content is solely the responsibility of the authors and does not necessarily represent the official views of the National Institute of General Medical Sciences or the National Institutes of Health. The authors declare no known conflict of interest.


\bibliography{bibs_v2}

\onecolumngrid
\appendix

\section{MAXIMUM ENTROPY DISTRIBUTIONS WITH CONSERVED MARGINALS}
\subsection{Two dimensions}
\label{MaxEnt_2D}
In two dimensions, we perform entropy maximization with the conservation of $x$ and $y$ marginal distributions considered with Lagrange multipliers:
\beq
\mathcal{S}[\rho] &= -\int \dd{x} \int \dd{y} \rho(x,y) \log \frac{\rho(x,y)}{q} + \lambda\left(1 - \int \dd{x} \int \dd{y} \rho\right) \\
&+ \int \dd{x} \phi_x(x) \left(\rho(x) - \int \dd{y} \rho(x,y)\right) + \int \dd{y} \phi_y(y) \left(\rho(y) - \int \dd{x} \rho(x,y)\right).
\eeq
Taking the functional derivative, we have
\beq
\fdv{\mathcal{S}}{\rho} = -\log \frac{\rho(x,y)}{q} - 1 - \lambda - \phi_x(x) - \phi_y(y) = 0,
\eeq
which gives us
\beq
\rho(x,y) = e^{ \ln q - 1 - \lambda - \phi_x(x) - \phi_y(y)}.
\eeq
Applying the normalization constraint, we have
\beq
\rho(x,y) = \frac{e^{- \phi_x(x) - \phi_y(y)}}{\int \dd{x} e^{- \phi_x(x)} \int \dd{y} e^{- \phi_y(y)}}.
\eeq
We now use the constraint on the marginal distribution over $x$, by integrating over $y$
\beq
\rho(x) = \int \dd{y} \rho(x,y) = \frac{e^{- \phi_x(x)} \int \dd{y} e^{- \phi_y(y)}}{\int \dd{x} e^{- \phi_x(x)} \int \dd{y} e^{- \phi_y(y)}} = \frac{e^{- \phi_x(x)}}{\int \dd{x} e^{- \phi_x(x)}},
\eeq
which allows us to replace $\phi_x(x)$. Repeating the same for $y$ gives us
\beq
\rho(y) = \frac{e^{- \phi_y(y)}}{\int \dd{y} e^{- \phi_y(y)}}.
\eeq
Thus, the final answer is
\beq
\rho(x,y) = \rho(x)\rho(y),
\eeq
which is the product of the one-dimensional marginal distributions.

\subsection{Arbitrary dimensions}
\label{MaxEnt_general}
In $d$ dimensions, suppose we have an initial distribution $\rho(\vb{r})$, where $\vb{r}$ is a position vector. As with the previous case, we perform entropy maximization with Lagrangian constraints on all $(k-1)$-dimensional marginal distributions:
\beq
\mathcal{S}[\rho] &= - \int \dd[d]{\vb{r}} \rho(\vb{r}) \log \rho(\vb{r}) +  \lambda\left(1 - \int \dd{x} \int \dd{y} \rho\right) \\
&+ \sum_{\substack{S\subseteq \{1,\dots,d\}\\\abs{S}=k-1}} \int \dd[k-1]{\vb{r}_S} \phi_S(\vb{r}_S) \left(\rho(\vb{r}_S) - \int \dd[d-k+1]{\vb{r}_{\{1,\dots,d\}\setminus S}} \rho(\vb{r})\right).
\eeq
Taking the functional derivative, we have
\beq
\fdv{\mathcal{S}}{\rho} = -\log \rho(\vb{r}) - 1 - \lambda - \sum_{\substack{S\subseteq \{1,\dots,d\}\\\abs{S}=k-1}} \phi_S(\vb{r}_S) = 0,
\eeq
which gives us
\beq
\rho(\vb{r}) = \exp\left[-1-\lambda - \sum_{\substack{S\subseteq \{1,\dots,d\}\\\abs{S}=k-1}} \phi_S(\vb{r}_S) \right].
\eeq
Applying the normalization constraint, we have
\beq
\rho^{\mathrm{eq}} = \frac{1}{Z}\exp\left[\sum_{\substack{S\subseteq \{1,\dots,d\}\\\abs{S}=k-1}} \phi_{S}(\vb{r}_S)\right],
\eeq
where
\beq
Z = \int \dd[d]{\vb{r}} \exp\left[\sum_{\substack{S\subseteq \{1,\dots,d\}\\\abs{S}=k-1}} \phi_{S}(\vb{r}_S)\right].
\eeq

\section{ENTROPY, CROSS-ENTROPY, AND KL DIVERGENCE IN MARGINAL-CONSERVING DIFFUSION}
\subsection{Monotonicity of Shannon entropy}
\label{lyapunov_entropy}
Following the approach of Ref.~\cite{han_scaling_2024}, for the general dynamics in \Cref{eq:gendiff}, we calculate the time derivative of the entropy
\beq
\dv{\mathcal{S}}{t} &= - \int \dd[d]{\vb{r}} \left[1 + \log \rho\right] \partial_t \rho \\
&=  (-1)^{k} D\int \dd[d]{\vb{r}} \left[1 + \log \rho\right]\sum_{\substack{S\subseteq \{1,\dots,d\}\\\abs{S}=k}} \left(\prod_{s \in S}\partial_{s}\right)\rho^{2^{k-1}} \left(\prod_{s \in S}\partial_{s}\right)\log \rho \\
&= (-1)^{k} D\sum_{\substack{S\subseteq \{1,\dots,d\}\\\abs{S}=k}}\int \dd[d]{\vb{r}} (-1)^{k} \left[\left(\prod_{s \in S}\partial_{s}\right)\left[1 + \log \rho\right]\right] \rho^{2^{k-1}} \left(\prod_{s \in S}\partial_{s}\right)\log \rho,
\eeq
where in the last step the alternating signs emerge from integration by parts with vanishing boundary terms. We thus obtain
\beq
\dv{\mathcal{S}}{t} &=  D\sum_{\substack{S\subseteq \{1,\dots,d\}\\\abs{S}=k}}\int \dd[d]{\vb{r}}  \left[\left(\prod_{s \in S}\partial_{s}\right)\left[\log \rho\right]\right]^2 \rho^{2^{k-1}} \geq 0,
\eeq
since all terms in the integrand are non-negative.

\subsection{KL divergence between initial and equilibrium distributions equals the entropy gap}
\label{lyapunov_kldiv}
The KL divergence between the particle density distribution and the equilibrium distribution is
\beq
D_{\mathrm{KL}}[\rho||\rho^{\mathrm{eq}}] &= \int \dd[d]{\vb{r}} \rho(\vb{r}) \log \frac{\rho(\vb{r})}{\rho^{\mathrm{eq}}(\vb{r})} \\
&= \int \dd[d]{\vb{r}} \rho(\vb{r}) \log \rho(\vb{r}) - \int \dd[d]{\vb{r}} \rho(\vb{r}) \log \rho^{\mathrm{eq}}(\vb{r}) \\
&= -\mathcal{S}[\rho] - \int \dd[d]{\vb{r}} \rho(\vb{r}) \log \rho^{\mathrm{eq}}(\vb{r}) \\
&= -\mathcal{S}[\rho] + \mathcal{S}[\rho|\rho^{\mathrm{eq}}],
\eeq
where $\mathcal{S}[\rho|\rho^{\mathrm{eq}}]$ is the cross-entropy between $\rho$ and $\rho^{\mathrm{eq}}$. We now expand the equilibrium distribution using the analytical form of the maximum-entropy distribution derived in Appendix \ref{MaxEnt_general}:
\beq
\mathcal{S}[\rho|\rho^{\mathrm{eq}}] &= - \int \dd[d]{\vb{r}} \rho(\vb{r}) \log \rho^{\mathrm{eq}}(\vb{r}) \\
&= - \int  \dd[d]{\vb{r}} \rho(\vb{r}) \left[\log \exp\left(\sum_{\substack{S\subseteq \{1,\dots,d\}\\\abs{S}=k-1}} \phi_{S}(\vb{r}_S)\right)\right] + \int \dd[d]{\vb{r}} \rho(\vb{r}) \log Z \\
&= - \sum_{\substack{S\subseteq \{1,\dots,d\}\\\abs{S}=k-1}}  \int  \dd[d]{\vb{r}} \rho(\vb{r}) \phi_{S}(\vb{r}_S) + \log Z \\
&= - \sum_{\substack{S\subseteq \{1,\dots,d\}\\\abs{S}=k-1}}  \int  \dd[k-1]{\vb{r}_S} \rho(\vb{r}_S) \phi_{S}(\vb{r}_S) + \log Z.
\eeq
Noting that the $(k-1)$-dimensional marginal distributions are conserved by the dynamics,
\beq
\rho(\vb{r}_S) = \rho^{\mathrm{eq}}(\vb{r}_S),
\eeq
we can rewrite
\beq
\mathcal{S}[\rho|\rho^{\mathrm{eq}}] &= - \sum_{\substack{S\subseteq \{1,\dots,d\}\\\abs{S}=k-1}}  \int  \dd[k-1]{\vb{r}_S} \rho^{\mathrm{eq}}(\vb{r}_S)\phi_{S}(\vb{r}_S) + \log Z \\
&= - \int  \dd[d]{\vb{r}} \rho^{\mathrm{eq}}(\vb{r}) \left[\log \exp\left(\sum_{\substack{S\subseteq \{1,\dots,d\}\\\abs{S}=k-1}} \phi_{S}(\vb{r}_S)\right)\right] + \int \dd[d]{\vb{r}} \rho^{\mathrm{eq}}(\vb{r}) \log Z \\
&= - \int \dd[d]{\vb{r}} \rho^{\mathrm{eq}}(\vb{r}) \log \rho^{\mathrm{eq}}(\vb{r}) \\
&= \mathcal{S}[\rho^{\mathrm{eq}}],
\eeq
which shows that the cross-entropy depends only on the equilibrium distribution.

\section{THREE-DIMENSIONAL GAUSSIAN WITH MONOTONICALLY DECREASING TOTAL CORRELATION AND NON-MONOTONIC MUTUAL INFORMATION}
\label{gaussian_appendix}
We want to find a three-dimensional Gaussian distribution which with a covariance matrix given in \Cref{eq:covariance_mat} whose time evolution under the diffusion process leads to \textit{increasing} mutual information between the $X(t)$ and $Y(t)$ coordinates while the total correlation between the three spatial variables monotonically decreases towards zero. First, we note that
\beq
I[X(t),Y(t)] = -\mathcal{S}[\rho(x,y,t)] + \mathcal{S}[\rho(x)] + \mathcal{S}[\rho(y)],
\eeq
where the first term is the negative entropy of the time-dependent two-dimensional marginal, and the second and third terms are entropies of time-independent one-dimensional marginals. Thus, our non-monotonicity condition \Cref{eq:nonmonotonicity_condition} becomes
\beq
\dv{\mathcal{S}[\rho(x,y,t)]}{t} < 0. 
\eeq
Writing the time-derivative of the entropy of the two-dimensional marginal exactly, we have
\beq
\dv{\mathcal{S}[\rho(x,y,t)]}{t} &= - \int \dd{x} \int \dd{y} \partial_t \left[\rho(x,y,t) \log \rho(x,y,t)\right] \\
&= - \int \dd{x} \int \dd{y} \left[\partial_t\rho(x,y,t)\right] \left[1 + \log \rho(x,y,t)\right]
\label{eq:timederivative_entropy_marginal}
\eeq
The hydrodynamic equations with the subsystem symmetries of mass density conserved along planes orthogonal to any Cartesian axis (i.e. one-dimensional marginal distribution-conserving diffusion) are given by \Cref{eq:gendiff}, which, for $d = 3$ and $k = 2$, are
\beq
\partial_t \rho(x,y,z,t) = - \partial_x\partial_y J_{xy} - \partial_y\partial_z J_{yz} - \partial_x\partial_z J_{xz},
\eeq
with currents given by
\beq
J_{ij} = D\rho^{2} \partial_{i} \partial_j \log \rho.
\eeq
Assuming that currents have vanishing boundary conditions, we can compute the time derivative of the two-dimensional marginal distribution
\beq
\partial_t \rho(x,y,t) &=  \int \dd{z} \partial_t \rho(x,y,z,t) \\
&= - \int \dd{z} \partial_x\partial_y J_{xy} - \partial_y\int \dd{z} \partial_z J_{yz} - \partial_x\int \dd{z}\partial_z J_{xz}, \\
&= - \partial_x\partial_y \int \dd{z} J_{xy} \\
&= - \partial_x\partial_y D \int \dd{z}  \rho(x,y,z,t)^{2} \partial_x \partial_y \log \rho(x,y,z,t).
\eeq
We then substitute this into \Cref{eq:timederivative_entropy_marginal}, integrate by parts with respect to $x$ and $y$, and simplify, assuming that boundary terms vanish:
\beq
\dv{\mathcal{S}[\rho(x,y,t)]}{t} &=  D \int \dd{x} \int \dd{y} \left[\int \dd{z}  \rho(x,y,z,t)^{2} \partial_x \partial_y \log \rho(x,y,z,t)\right] \left[1 + \log \rho(x,y,t)\right] \\
&= D \int \dd{x} \int \dd{y} \left[ \int \dd{z}  \rho(x,y,z,t)^{2} \partial_x \partial_y \log \rho(x,y,z,t)\right] \partial_x\partial_y \left[1 + \log \rho(x,y,t)\right] \\
&= D \int \dd{x} \int \dd{y} \left[ \int \dd{z}  \rho(x,y,z,t)^{2} \partial_x \partial_y \log \rho(x,y,z,t)\right] \partial_x\partial_y  \log \rho(x,y,t) \\
&= D \int \dd{x} \int \dd{y} \int \dd{z}  \rho(x,y,z,t)^{2} \left[\partial_x \partial_y \log \rho(x,y,z,t)\right]\left[\partial_x\partial_y  \log \rho(x,y,t)\right].
\eeq
It now follows that the mutual information's non-monotonicity condition is met only when the product
\beq
\left[\partial_x \partial_y \log \rho(x,y,z,t)\right]\left[\partial_x\partial_y  \log \rho(x,y,t)\right] < 0.
\eeq

The PDF of the original Gaussian distribution at $t= 0$ is (dropping the explicit $t$ dependence):
\beq
\rho(x,y,z) = \frac{1}{\sqrt{(2\pi)^3 \det \Sigma}} \exp\left[-\frac{1}{2} \vb{r}^T \Sigma^{-1} \vb{r}\right],
\eeq
where $\vb{r} = (x,y,z)$, and $\Sigma^{-1}$ is the inverse of the covariance matrix from in \Cref{eq:covariance_mat}, given by
\beq
\Sigma^{-1} = \frac{1}{\det \Sigma}\begin{pmatrix}
1-\sigma_{yz}^2 & \sigma_{xz} \sigma_{yz}-\sigma_{xy} & \sigma_{xy} \sigma_{yz}-\sigma_{xz} \\
 \sigma_{xz} \sigma_{yz}-\sigma_{xy} & 1-\sigma_{xz}^2 & \sigma_{xy} \sigma_{xz}-\sigma_{yz} \\
 \sigma_{xy} \sigma_{yz}-\sigma_{xz} & \sigma_{xy} \sigma_{xz}-\sigma_{yz} & 1-\sigma_{xy}^2
\end{pmatrix},
\eeq
with
\beq
\det \Sigma = 1 - \sigma_{xy}^2 - \sigma_{xz}^2 - \sigma_{yz}^2 + 2 \sigma_{xy} \sigma_{xz} \sigma_{yz}.
\eeq
The marginal distribution obtained from integrating over $z$ is
\beq
\rho(x,y) = \frac{1}{\sqrt{(2\pi)^2 \det \Xi}} \exp\left[-\frac{1}{2} \vb{s}^T \Xi^{-1} \vb{s}\right],
\eeq
where $\vb{s} = (x,y)$, and 
\beq
\Xi = \begin{pmatrix}
1 & \sigma_{xy} \\
\sigma_{xy} & 1
\end{pmatrix}.
\eeq
The inverse of $\Xi$ is
\beq
\Xi^{-1} = \frac{1}{\det \Xi}\begin{pmatrix}
1 & -\sigma_{xy} \\
-\sigma_{xy} & 1
\end{pmatrix},
\eeq
where
\beq
\det \Xi = 1 - \sigma_{xy}^2.
\eeq

The first term in the non-monotonicity condition is
\beq
\partial_x \partial_y \log \rho(x,y,z) &= \partial_x \partial_y \left[-\frac{1}{2} \vb{r}^T \Sigma^{-1} \vb{r} - \frac{1}{2}\log \left((2\pi)^3 \det \Sigma\right)\right] \\
&= \frac{\sigma_{xy} - \sigma_{xz}\sigma_{yz}}{\det \Sigma},
\eeq
and the second term is
\beq
\partial_x\partial_y  \log \rho(x,y) &= \partial_x \partial_y \left[-\frac{1}{2} \vb{s}^T \Xi^{-1} \vb{s} - \frac{1}{2}\log \left((2\pi)^2 \det \Xi\right)\right] \\
&= \frac{\sigma_{xy}}{\det \Xi}.
\eeq
Thus, the non-monotonicity condition becomes
\beq
\frac{\sigma_{xy} - \sigma_{xz}\sigma_{yz}}{\det \Sigma} \frac{\sigma_{xy}}{\det \Xi} < 0.
\eeq
Since the determinants are positive for the positive-definite covariance matrices, we have a simplified condition
\beq
\sigma_{xy}(\sigma_{xy} - \sigma_{xz}\sigma_{yz})  < 0,
\eeq
which is \Cref{eq:final_condition}. Note that the correlations must be chosen so that the covariance matrix remains positive semi-definite. An example set of values is
\beq
\sigma_{xy} = 0.1, \quad \sigma_{xz} = 0.5, \quad \sigma_{yz} = 0.5.
\eeq

\section{Derivation of the equilibrium noise correlation \Cref{eq:2d_fluc}} \label{appen:noise}

In this section, we derive the statistics of the noise correlation~\Cref{eq:2d_fluc} by using the fact that the system reaches equilibrium and therefore does not produce entropy in the steady state. To start, we rewrite the governing equations, Eqs.~\eqref{eq:2d_shear} and \eqref{eq:2d_current}, as 
\begin{align}
    \nonumber \partial_t \rho({\bf r}, t) &= - \partial_i \partial_j J_{ij} ({\bf r}, t) \\
    J_{ij}({\bf r}, t) &= \mu_{ijkl} \partial_k \partial_l \frac{\delta {\cal F}}{\delta \rho} + \xi_{ij}({\bf r}, t) \, , \label{eq:J_FreeEnergy}
\end{align}
where the free energy ${\cal F}[\rho]$ given by
\begin{align}
    {\cal F}[\rho] = k_{\rm B} T \int {\rm d}{\bf r} \, \rho({\bf r}, t) \ln \frac{\rho({\bf r}, t)}{q} = -T S[\rho] \, ,
    \label{eq:cal_F}
\end{align}
with Boltzmann constant $k_{\rm B}$, temperature $T$, and entropy $S$. The noise term $\xi_{ij}$ has zero mean and correlation
\begin{align}
    \langle \xi_{ij}({\bf r}, t) \xi_{kl}({\bf r}', t') \rangle
    = 2D_{ijkl} \delta({\bf r} - {\bf r}') \delta(t - t') \, ,
\end{align}
where we assume that the noise is local in space and time, so that its correlation can be approximated by delta distributions. 

Next, we approximate $\xi_{ij}$ as a Gaussian noise. Its PDF then satisfies 
\begin{align} \nonumber
    {\cal P}[\xi] &\propto \exp \left[ - \frac{1}{4} \int {\rm d} {\bf r} \, {\rm d}t \int {\rm d} {\bf r}' \, {\rm d} t' \, \xi_{ij} ({\bf r}, t) D^{-1}_{ijkl} \delta({\bf r} - {\bf r}') \delta( t - t') \xi_{kl}({\bf r}', t') \right] \\
    &= \exp \left[ - \frac{1}{4} \int {\rm d} {\bf r} \, {\rm d} t \, \xi_{ij} ({\bf r}, t) D_{ijkl}^{-1} \xi_{kl}({\bf r}, t) \right].
\end{align}
We now convert ${\cal P}[\xi]$ into the PDF for the density field. This leads to~\cite{martin1973statistical}
\begin{align} \nonumber
    {\cal P}[\rho] \propto \frac{{\cal D}[\xi]}{{\cal D}[\rho]} \exp \left[ - \frac{1}{4} \int {\rm d} {\bf r} {\rm d}t \, \left\{ J_{ij}({\bf r}, t) - \mu_{ijmn} \partial_m \partial_n \frac{\delta {\cal F}}{\delta \rho({\bf r}, t)} \right\} D_{ijkl}^{-1} \left\{ J_{kl}({\bf r}, t) - \mu_{klpq} \partial_p \partial_q \frac{\delta {\cal F}}{\delta \rho({\bf r}, t)} \right\} \right] \, ,
\end{align}
with Jacobian ${\cal D}[\xi]/{\cal D}[\rho]$. We compute the entropy production from the Kullback-Leibler divergence between the probabilities of observing a forward trajectory and its time-reversed counterpart:
\begin{align} \label{eq:S}
    {\cal S} = k_{\rm B} \left\langle \ln\frac{ {\cal P}[\rho]}{{\cal P}^{\rm R}[\rho]} \right\rangle \, .
\end{align}
Here ${\cal P}^{\rm R}[\rho] = {\cal P}[\tilde{\rho}]$ denotes the probability of the time-reversed trajectory $\tilde{\rho}({\bf r}, t) \equiv \rho({\bf r}, t_{\rm f} - t)$, where $t_{\rm f}$ is the final observation time. Inserting this relation into Eq.~\eqref{eq:J_FreeEnergy}, we obtain
\begin{align}
    \partial_t \tilde{\rho}({\bf r}, t) = \partial_t \rho({\bf r}, t_{\rm f} -  t) = -\partial_\tau \rho({\bf r}, \tau)|_{\tau = t_{\rm f} - t} =  \partial_i \partial_j  J_{ij} ({\bf r}, t_{\rm f} - t )
\end{align}
which lets us conclude that the current is a time-reversal odd variable, i.e., $\tilde{J}_{ij}({\bf r}, t) = - J_{ij}({\bf r}, t_{\rm f} - t)$.

Using the time-reversal properties of $\rho$ and $J_{ij}$, we can rewrite Eq.~\eqref{eq:S} as 
\begin{align} \nonumber
    {\cal S} [\rho] &= k_{\rm B} \left\langle \ln \frac{{\cal P}[\rho]}{{\cal P}[\tilde{\rho}]}
    \right\rangle \\
    &= k_{\rm B} \left\langle \int {\rm d} {\bf r} {\rm d} t \, \left[ J_{ij} D_{ijkl}^{-1} \mu_{klpq} \partial_p \partial_q \frac{\delta {\cal F}}{\delta \rho({\bf r}, t)}  \right]\right\rangle \, .
\end{align}
Now, if we set 
\begin{align} \label{eq:D_and_mu}
    D_{ijkl}^{-1} \mu_{klpq} = A \delta_{ip} \delta_{jq} \, ,
\end{align} 
then the entropy production becomes
\begin{align} \nonumber
    {\cal S} &= k_{\rm B} A\left\langle \int {\rm d} {\bf r} {\rm d} t \, J_{ij} \partial_i \partial_j \frac{\delta {\cal F}}{\delta \rho({\bf r}, t)} \right\rangle \\
    &= k_{\rm B} A\left\langle  \int {\rm d} {\bf r} {\rm d} t \,  (\partial_i \partial_j J_{ij} ) \frac{\delta {\cal F}}{\delta \rho({\bf r}, t)}  \right\rangle \\
    &= - k_{\rm B} A \left\langle \int {\rm d} {\bf r} {\rm d} t \, \partial_t \rho({\bf r}, t)   \frac{\delta {\cal F}}{\delta \rho({\bf r}, t) } \right\rangle \\
    &= - k_{\rm B} A \left\langle \int {\rm d} {\bf r} {\rm d} t \, \frac{{\rm d}}{{\rm d} t}   { {\cal F}} \right\rangle \\
    &= k_{\rm B} A\left[  \int {\rm d} {\bf r} \left\langle {\cal F}[\rho({\bf r}, 0)] \right\rangle 
    - \int {\rm d} {\bf r} \left\langle {\cal F}[\rho({\bf r}, t_{\bf f} )] \right\rangle \right] \, ,
\end{align}
where we performed an integral by parts in going from the first line to the second. Using the definition of the free energy, the entropy production becomes
\begin{align}
    {\cal S} = Ak_{\rm B}T \langle S(t_{\bf f}) - S(0) \rangle \, .
\end{align}
Therefore, the relation $A = (k_{\rm B} T)^{-1}$ should hold for the entropy production to match the entropy change in the system. Substituting this relation into \Cref{eq:D_and_mu}, we obtain the generalized Einstein relation
\begin{align}
    D_{ijkl} = \mu_{ijkl} k_{\rm B} T \, .
\end{align}
This shows that \Cref{eq:2d_fluc} is consistent with the second law of thermodynamics and can therefore serve as the noise correlation for systems that relax to equilibrium.

\end{document}